\documentclass[aps,pra,floatfix,twocolumn,showpacs]{revtex4}

\usepackage[dvipdfm,pdfstartview=FitH]{hyperref}
\usepackage{float}
\usepackage{graphicx}
\usepackage[english]{babel}
\usepackage{amsmath}
\usepackage{amssymb}
\usepackage{slashed}
\allowdisplaybreaks
\begin{document}
\title{Gauge Invariant Linear Response Theories for Ultracold Fermi Gases with Pseudogap}

\author{Hao Guo$^{1}$, Yan He$^{2}$}
\affiliation{$^1$Department of Physics, Southeast University, Nanjing 211189, China}
\affiliation{$^2$College of Physical Science and Technology, Sichuan University, Chengdu, Sichuan 610064, China}
\begin{abstract}
Recent experimental progresses allow for exploring some important physical quantities of ultracold Fermi gases, such as the compressibility, spin susceptibility, viscosity, optical conductivity and spin diffusivity. Theoretically, these quantities can be evaluated from suitable linear response theories. For BCS superfluid, it has been found that the gauge invariant linear response theories can be fully consistent with some stringent consistency constraints. When the theory is generalized to stronger-than-BCS regime, one may meet serious difficulties to satisfy the gauge invariance conditions. In this paper, we try to construct density and spin linear response theories which are formally gauge invariant for a Fermi gas undergoing BCS-Bose-Einstein Condensation (BEC) crossover, especially below the superfluid transition temperature $T_c$. We adapt a particular $t$-matrix approach which is close to the $G_0G$ formalism to incorporate non-condensed pairing in the normal state. We explicitly show that the fundamental constraints imposed by the Ward identities, $Q$-limit Ward identity are indeed satisfied.
\end{abstract}


\maketitle

\section{Introduction}
Recently there is a broad literature on the subjects of response functions in superconductors and atomic Fermi gas superfluids \cite{Arseev,Stringari06,HaussmannPRL12,StrinatiPRL12,HaoPRL10,HHEPL10,HLPRA10,HLPRA12,HLPRA10-2,SadeMeloC11,YanPRB14}, where inter-particle interaction is strong enough such that the classical BCS theory is not adequate here. Related experiments include the studies of the thermodynamic response functions and dynamical response\cite{Ketterle11,ZwierleinNature11,ValePRL,ValePRL12, Zwierlein12}. Theoretically, linear response theories have been an important tool for studying the transport and dynamic properties of Fermi gases. Hence, it is important to assess the self consistency of the linear response theories as well as comparing the experimental results. There must be some general rules that the theory must follow. In the references \cite{OurJLTP13,HaoJPB14}, several fundamental constraints associated with the conservation laws/Ward identities and sum rules were addressed. Since the conservation laws are generically related to some (gauge) symmetry of the theory, then in the broken-symmetry phase or ordered phase it is particular difficult for many-body theories to satisfy all these constraints. It was also pointed out that the strict weakly interacting BCS mean field theory does pass all these testings both below and above $T_c$ even when the pairing population is unbalanced \cite{OurJLTP13,HaoJPB14}. In other words, the linear response theories of BCS superfluids can be formulated into a fully gauge invariant theory. In the normal state, the simplest Nozieres Schimitt-Rink (NSR) is also compatible with these gauge invariance condition \cite{OurIJMPB}. However, it is well known that the BCS mean field theory is not suitable to describe the Fermionic superfluid when the inter-particle interaction becomes strong. Moreover, in the broken-symmetry phase, the consistent generalization of the linear response theory based on the NSR theory may meet great difficulties since a first order transition appears at the symmetry-breaking temperature, $T_c$.

In this paper, we try to build an ideal linear response theory for strongly correlated superdfluids undergoing BCS-BEC crossover by a diagrammatic approach such that the fundamental constraints mentioned above can be satisfied. Our selected diagrams bear on those associated with the Goldstone modes due to the symmetry-breaking via the consistent-fluctuation-of the order parameter (CFOP) approach, and conventional contributions, namely, the Maki-Thompson (MT) and Aslamazov-Larkin (AL) diagrams. As a price, we have to adapt a slightly modified $G_0G$ formalism to incorporate the pairing fluctuation effect. We emphasize that this approach is purely a theoretical attempt until now. However, it might be a necessary step to fully understand the transport properties of strongly correlated Fermi gases.

The linear response theories must be consistent with several fundamental constraints \cite{Mahanbook,OurJLTP13,HaoJPB14} imposed by the Ward identities, $Q$-limit Ward identity \cite{Maebashi09}. It is well known that the Ward identities guarantees the gauge invariance of the theory, while the $Q$-limit Ward identities lead to the sum rules of compressibility and spin susceptibilities which further build the consistent connection between the single-particle thermodynamics and two-particle correlation functions. However, for spin response theory the $Q$-limit Ward identity is only meaningful to polarized Fermi superfluids \cite{HaoJPB14}. In this paper, we focus on the unpolarized Fermi superfluids. The central difficulty of formulating the consistent linear response theory is to maintain the gauge invariance when the pseudogap self-energy is introduced by the pairing fluctuation effect. This is obviously beyond the CFOP approach since the total energy gap is now different from the order parameter.

In the following sections, we first briefly review the CFOP linear response theories both in the density and spin channels for BCS mean field theory, then we introduce the pairing fluctuation effects via a particular $t$-matrix formalism. We further carry on extra diagrammatic corrections in the two channels, and verify that the new theories do maintain the gauge symmetry respectively. Throughout this paper, we follow the convention $c=\hbar=k_B=1$.

\section{BCS mean field theory approach}
By using the $\sigma$ to denote the spin or pseudo-spin $\uparrow,\downarrow$, the Hamiltonian for a two component Fermi gas interacting via the attractive contact interaction $g$ is
\begin{eqnarray}\label{H00}
H&=&\int d^3\mathbf{x}\psi^{\dagger}_{\sigma}(\mathbf{x})\Big(\frac{\hat{\mathbf{p}}^2}{2m}-\mu\Big)\psi_{\sigma}(\mathbf{x})\nonumber\\
& &-g\int
d^3\mathbf{\mathbf{x}}\psi^{\dagger}_{\uparrow}(\mathbf{x})\psi^{\dagger}_{\downarrow}(\mathbf{x})\psi_{\downarrow}(\mathbf{x})\psi_{\uparrow}(\mathbf{x}),
\end{eqnarray}
where $\psi$ and $\psi^{\dagger}$ are the annihilation and creation operators of fermions, $\mu$ is the chemical potential and $m$ is the fermion mass. There is an implicit summation over the pseudo-spin indices $\sigma$. The Hamiltonian has a U(1)$\times$ U(1) symmetry \cite{Nambu60}
\begin{eqnarray}\label{T2}
& &\psi_{\sigma}\rightarrow e^{-i\alpha}\psi_{\sigma},\quad \psi^{\dagger}_{\sigma}\rightarrow e^{i\alpha}\psi^{\dagger}_{\sigma}; \nonumber \\
& &\psi_{\sigma}\rightarrow e^{-iS_{\sigma}\alpha}\psi_{\sigma},\quad \psi^{\dagger}_{\sigma}\rightarrow e^{iS_{\sigma}\alpha}\psi^{\dagger}_{\sigma}, \label{sz}
\end{eqnarray}
where $S_{\uparrow,\downarrow}=\pm 1$ and $\alpha$ is the phase parameter of those transformations. The first U(1) symmetry is well known for relating to the electromagnetism (EM). If the particle is charged, this symmetry naturally becomes a gauge symmetry. For a charge neutral system, the symmetry is still associated with the mass current conservation. The second symmetry is the spin rotational symmetry which is associated with the spin current conservation. Our linear response theories in the density and spin channels must respect these two symmetries respectively. The central idea is to ``gauge'' the U(1) symmetries by introducing two types of weak external fields. In the density channel, it is the weak EM field $A^{\mu}=(\phi,\mathbf{A})$. While in the spin channel, it is $A^{\mu}\equiv(B_z,\mathbf{m})$, where $B_z$ is the $z$ component of the magnetic field (assuming $z$ is the axis of spin rotation) and $\mathbf{m}$ is the magnetization.

After taking BCS mean field approximation, the order parameter or superconducting gap function is introduced $\Delta_{\textrm{sc}}(\mathbf{x})=-g\langle\psi_{\uparrow}(\mathbf{x})\psi_{\downarrow}(\mathbf{x})\rangle$. The first U(1) symmetry is spontaneously broken while the second is not. As can be seen in the reference \cite{OurJLTP13}, this brings significant difference between the linear response theories in the two channels. For a homogeneous system, the BCS Hamiltonian can be expressed as
\begin{eqnarray}\label{HBCS}
H_{\textrm{BCS}}&=&\sum_{\mathbf{k}\sigma}\psi^{\dagger}_{\mathbf{p}\sigma}\xi_{\mathbf{p}}\psi_{\mathbf{p}\sigma}\nonumber\\
&+&\sum_{\mathbf{p}}\Delta_{\textrm{sc}}\psi_{-\mathbf{p}\uparrow}\psi_{\mathbf{p}\downarrow}
+\sum_{\mathbf{p}}\Delta_{\textrm{sc}}\psi^{\dagger}_{\mathbf{p}\downarrow}\psi^{\dagger}_{-\mathbf{p}\uparrow},
\end{eqnarray}
where $\xi_{\mathbf{p}}=\frac{\mathbf{p}^2}{2m}-\mu$. As a familiar result, the BCS Green and anomalous Green functions in the momentum space are
\begin{eqnarray}\label{GFE2}
& &G_{\textrm{sc}}(i\omega_n,\mathbf{p})=\frac{u^2_{\textrm{sc}\mathbf{p}}}{i\omega_n-E_{\textrm{sc}\mathbf{p}}}+\frac{v^2_{\textrm{sc}\mathbf{p}}}{i\omega_n+E_{\textrm{sc}\mathbf{p}}},\nonumber\\
& &F_{\textrm{sc}}(i\omega_n,\mathbf{p})=-\frac{u_{\textrm{sc}\mathbf{p}}v_{\textrm{sc}\mathbf{p}}}{i\omega_n-E_{\textrm{sc}\mathbf{p}}}+\frac{u_{\textrm{sc}\mathbf{p}}v_{\textrm{sc}\mathbf{p}}}{i\omega_n+E_{\textrm{sc}\mathbf{p}}}.
\end{eqnarray}
where $i\omega_n$ is the Fermion Matsubara frequency, and $E_{\textrm{sc}\mathbf{p}}=\sqrt{\xi^2_{\mathbf{p}}+\Delta^2_{\textrm{sc}}}$ is the quasi-particle energy dispersion. Hereinafter we use the subscript ``sc'' to emphasize that these discussions are only under the BCS mean field approximation. Define $P\equiv (i\omega_n,\mathbf{p})$, the number and gap equations are determined by
$n=2\sum_PG_{\textrm{sc}}(P)$ and $\Delta_{\textrm{sc}}=-g\sum_PF_{\textrm{sc}}(P)$. These identities give
\begin{eqnarray}
n&=&\sum_{\mathbf{p}}\big[1-\frac{\xi_{\textrm{sc}\mathbf{p}}}{E_{\textrm{sc}\mathbf{p}}}(1-2f(E_{\textrm{sc}\mathbf{p}}))\big],\nonumber\\
\frac{1}{g}&=&\sum_{\mathbf{p}}\frac{1-2f(E_{\textrm{sc}\mathbf{p}})}{2E_{\textrm{sc}\mathbf{p}}}.
\end{eqnarray}
The bare Green function is $G_0(P)=(i\omega_n-\xi_{\mathbf{p}})^{-1}$. The Dyson equation gives $G^{-1}_{\textrm{sc}}(P)=G^{-1}_0(P)-\Sigma_{\textrm{sc}}(P)$ where $\Sigma_{\textrm{sc}}(P)=-\Delta^2_{\textrm{sc}}G_0(-P)$ is the BCS self-energy.
\subsection{Density Channel}
\begin{figure}[H]
\begin{center}
\includegraphics[width=1.5in,clip]
{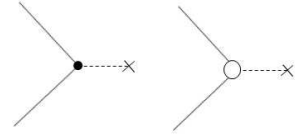}
\caption{EM interaction vertex. The left one is the bare vertex, and the right one is the full vertex. The solid line denotes the fermion line, the dashed line denotes the photon line.}
\label{fig:DV}
\end{center}
\end{figure}

In the density channel, the system is perturbed by an effective external EM field $A^{\mu}$, and the Hamiltonian becomes $H=H_{\textrm{BCS}}+H_{\textrm{I}}$ with
\begin{eqnarray}\label{H20}
H_{\textrm{I}}=\sum_{\mathbf{p}\mathbf{q}\sigma}\psi^{\dagger}_{\mathbf{p}+\mathbf{q}\sigma}\gamma^{\mu}(\mathbf{p}+\mathbf{q},\mathbf{p})A_{\mu\mathbf{q}}\psi_{\mathbf{p}\sigma}.
\end{eqnarray}
where $\gamma^{\mu}(\mathbf{p}+\mathbf{q},\mathbf{p})\equiv\gamma^{\mu}(P+Q,P)=(1,\frac{\mathbf{p}+\frac{\mathbf{q}}{2}}{m})$ is the bare EM interaction vertex. Here $Q\equiv q^{\mu}=(i\Omega_l,\mathbf{q})$ is the external four momentum, where $\Omega_l$ is the boson Matsubara frequency. The bare vertex satisfies the ``bare'' Ward identity
\begin{eqnarray}\label{BWI}
q_{\mu}\gamma^{\mu}(P+Q,P)=G^{-1}_{0}(P+Q)-G^{-1}_{0}(P).
\end{eqnarray}
In a gauge invariant EM linear response theory, a full EM interaction vertex $\Gamma^{\mu}$ (The bare and full EM vertices are shown in Figure.\ref{fig:DV}) which satisfies the full
Ward identity
\begin{eqnarray}\label{WI0}
q_{\mu}\Gamma^{\mu}_{\textrm{sc}}(P+Q,P)=G^{-1}_{\textrm{sc}}(P+Q)-G^{-1}_{\textrm{sc}}(P)
\end{eqnarray}
must be found, so that the perturbed current can be expressed as $\delta J^{\mu}(Q)=K^{\mu\nu}_{\textrm{sc}}(Q)A_{\nu}(Q)$ with $K^{\mu\nu}_{\textrm{sc}}(Q)$ determined by the Kubo formalism
\begin{eqnarray}
& &K^{\mu\nu}_{\textrm{sc}}(Q)=\frac{n}{m}h^{\mu\nu}\\
&+&2\sum_{P}\Gamma^{\mu}_{\textrm{sc}}(P+Q,P)G_{\textrm{sc}}(P+Q)\gamma^{\nu}(P,P+Q)G_{\textrm{sc}}(P),\nonumber
\end{eqnarray}
where $h^{\mu\nu}=-\eta^{\mu\nu}(1-\eta^{\nu0})$ with $\eta^{\mu\nu}=\textrm{diag}(1,-1,-1,-1)$ being the metric tensor. By using the Ward identity (\ref{WI0}), it's easy to show that $q_{\mu}K^{\mu\nu}_{\textrm{sc}}(Q)=0$, which further leads to the conservation of perturbed current $q_{\mu}\delta J^{\mu}(Q)=0$. Hence the linear response theory is indeed gauge invariant. Under the framework of BCS mean field theory, such full vertex can be obtained either by Nambu's integral-equation approach \cite{Nambu60,Schrieffer_book} or by the CFOP approach. However, the $Q$-limit Ward identity provides an independent consistency check of the theory. The vertex given by the latter approach is proved to satisfy this condition
\begin{eqnarray}\label{QWI}
\lim_{\mathbf{q}\rightarrow\mathbf{0}}\Gamma^0_{\textrm{sc}}(P+Q,P)|_{\omega=0}=\frac{\partial G^{-1}_{\textrm{sc}}(P)}{\partial \mu}=1-\frac{\partial \Sigma_{\textrm{sc}}(P)}{\partial \mu}.
\end{eqnarray}
Details can be found in the reference \cite{OurJLTP13}. This identity not only builds a consistent connection between the one-particle thermodynamics and two-particle response functions but also the acts as the sufficient and necessary condition for the compressibility sum rule
\begin{eqnarray}\label{CSR}
\frac{\partial n}{\partial \mu}=-K^{00}_{\textrm{sc}}(0,\mathbf{q}\rightarrow\mathbf{0}).
\end{eqnarray}
\begin{figure}[H]
\begin{center}
\includegraphics[width=0.6in,clip]
{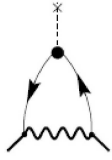}
\caption{Maki-Thompson diagram. The wavy line represents the pair propagator.}
\label{fig:MT}
\end{center}
\end{figure}
The expression of this gauge invariant interaction vertex given by the CFOP approach is
\begin{eqnarray}\label{V0}
\Gamma^{\mu}_{\textrm{sc}}(P+Q,P)&=&\gamma^{\mu}(P+Q,P)+\textrm{Coll}^{\mu}_{\textrm{sc}}(P+Q,P)\nonumber\\
&+&\textrm{MT}^{\mu}_{\textrm{sc}}(P+Q,P),
\end{eqnarray}
where the second term,
\begin{eqnarray}\label{C0}
& &\textrm{Coll}^{\mu}_{\textrm{sc}}(P+Q,P)\nonumber\\
&=&\Delta_{\textrm{sc}}\Pi^{\mu}(Q)G_0(-P-Q)+\Delta_{\textrm{sc}}\bar{\Pi}^{\mu}(Q)G_0(-P),
 \end{eqnarray}
  corresponds to the excitations of Nambu-Goldstone modes due to the breaking of the U(1) symmetry, and the third term,
\begin{eqnarray}\label{MTSC}
& &\textrm{MT}^{\mu}_{\textrm{sc}}(P+Q,P)\nonumber\\
&=&-\Delta^2_{\textrm{sc}}G_{0}(-P)\gamma^{\mu}(-P,-P-Q)G_0(-P-Q),
 \end{eqnarray}
 is the famous MT diagram which is shown in Figure.\ref{fig:MT}. The expressions of $\Pi^{\mu}$ and $\bar{\Pi}^{\mu}$ are given in Appendix.\ref{app1}. These two terms originates from summing up the diagrams with photon-fermion interaction lines inserted at any possible position.
 By using the equalities \cite{OurJLTP13} $q_{\mu}\Pi^{\mu}(Q)=2\Delta_{\textrm{sc}}$ and $q_{\mu}\bar{\Pi}^{\mu}(Q)=-2\Delta_{\textrm{sc}}$, it can be shown that
\begin{eqnarray}\label{CWI}
q_{\mu}\textrm{Coll}^{\mu}_{\textrm{sc}}(P+Q,P)=2\Sigma_{\textrm{sc}}(P)-2\Sigma_{\textrm{sc}}(P+Q).
\end{eqnarray}
Combining with the equality
\begin{eqnarray}\label{MWI}
q_{\mu}\textrm{MT}^{\mu}_{\textrm{sc}}(P+Q,P)=\Sigma_{\textrm{sc}}(P+Q)-\Sigma_{\textrm{sc}}(P),
\end{eqnarray}
the Ward identity (\ref{WI0}) in the BCS mean field level can be proved.

\subsection{Spin Channel}
\begin{figure}[H]
\begin{center}
\includegraphics[width=1.5in,clip]
{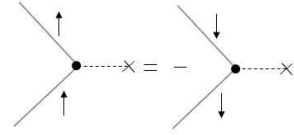}
\caption{Spin interaction vertex. It has different signs for different pseudo-spin indices.}
\label{fig:S-V}
\end{center}
\end{figure}
In the spin channel, the U(1) spin rotational symmetry is also ``gauged'' by introducing an effective external field. However, this symmetry is not broken by the order parameter. Therefore we expect that the structure of the spin linear response theory is simpler than that of its density counterpart. The ``bare'' spin interaction vertex is dependent on the pseudo-spin: $\gamma^{\mu}_{\textrm{S}\sigma}(\mathbf{p}+\mathbf{q},\mathbf{p})\equiv\gamma^{\mu}_{\textrm{S}\sigma}(P+Q,P)=S_{\sigma}(1,\frac{\mathbf{p}+\frac{\mathbf{q}}{2}}{m})$ with the subscript ``S'' referring to the ``spin''. The quantity $S_{\sigma}$ indicates that the vertex has different signs for different pseudo-spin indices, see Figure.\ref{fig:S-V}. It also respect the ``bare'' Ward identity in the spin channel
\begin{eqnarray}\label{BSWI}
q_{\mu}\gamma^{\mu}_{\textrm{S}\sigma}(P+Q,P)=S_{\sigma}\big(G^{-1}_{0}(P+Q)-G^{-1}_{0}(P)\big).
\end{eqnarray}
The spin interaction Hamiltonian is given by \begin{eqnarray}\label{HIS0}H_{\textrm{IS}}=\sum_{\mathbf{p}\mathbf{q}\sigma}\psi^{\dagger}_{\mathbf{p}+\mathbf{q}\sigma}\gamma^{\mu}_{\textrm{S}\sigma}(\mathbf{p}+\mathbf{q},\mathbf{p})A_{\mu\mathbf{q}}\psi_{\mathbf{p}\sigma}.\end{eqnarray}
Similarly, the perturbed spin current is also evaluated via the Kubo formalism $\delta J^{\mu}_{\textrm{S}}(Q)=K^{\mu\nu}_{\textrm{Ssc}}(Q)A_{\nu}(Q)$ where
\begin{eqnarray}
K^{\mu\nu}_{\textrm{Ssc}}(Q)&=&\frac{n}{m}h^{\mu\nu}+\sum_{P\sigma}\Gamma^{\mu}_{\textrm{Ssc}\sigma}(P+Q,P)\nonumber\\
&\times&G_{\textrm{sc}}(P+Q)\gamma^{\nu}_{\textrm{S}\sigma}(P,P+Q)G_{\textrm{sc}}(P).
\end{eqnarray}
The full spin interaction vertex $\Gamma^{\mu}_{\textrm{Ssc}\sigma}$ is given by
\begin{eqnarray}\label{SV0}
\Gamma^{\mu}_{\textrm{Ssc}\sigma}(P+Q,P)&=&\gamma^{\mu}_{\textrm{S}\sigma}(P+Q,P)\nonumber\\
&+&\textrm{MT}^{\mu}_{\textrm{Ssc}\sigma}(P+Q,P),
\end{eqnarray}
where the MT term is expressed as
\begin{eqnarray}\label{MTSSC}
& &\textrm{MT}^{\mu}_{\textrm{Ssc}\sigma}(P+Q,P)\nonumber\\
&=&-\Delta^2_{\textrm{sc}}G_{0}(-P)\gamma^{\mu}_{\textrm{S}\sigma}(-P,-P-Q)G_0(-P-Q).
 \end{eqnarray}
Since the U(1) spin rotational symmetry is not broken below $T_c$, then the full spin interaction vertex does not contain contributions associated with the Nambu-Goldstone modes. Moreover, the Ward identity is indeed satisfied in the mean field theory level
\begin{eqnarray}\label{SWI0}
q_{\mu}\Gamma^{\mu}_{\textrm{Ssc}\sigma}(P+Q,P)=S_{\sigma}\big(G^{-1}_{\textrm{sc}}(P+Q)-G^{-1}_{\textrm{sc}}(P)\big).
\end{eqnarray}
Therefore the perturbed spin current is conserved $q_{\mu}\delta J^{\mu}_{\textrm{S}}(Q)=0$. In the spin channel, there is no well-defined $Q$-limit Ward identity for unpolarized Fermi superfluids although such identity does exist for polarized Fermi superfluids \cite{HaoJPB14}. This is because the equal-population case can not be approached from the population imbalanced case by simply letting the particle number difference approach zero.

The above discussions show that the linear response theories in the density and spin channels are fully consistent with the BCS mean field approximation for Fermi superfluids. However, when generalized to the whole BCS-BEC crossover regime, the mean field approximation overestimates the critical temperature in the unitarity and BEC side since the fluctuations of the non-condensed pairs are ignored. We next show a formally theoretical scheme in which the fundamental constraints are still satisfied when the pairing fluctuation effects are included.

\section{Gauge invariant linear response theories in the $G_0G$ formalism}
When we consider the situation that the interaction between fermions is stronger than the BCS attraction, the self-energy obtains corrections from the non-condensed pairs. Hence the interaction vertex must be corrected correspondingly to ensure an exact validity of the Ward identity and $Q$-limit Ward identity. In this paper, we adapt the $G_0G$ formalism \cite{OurRoPP} to discuss the pseudo-gap effect. The self-energy due to the non-condensed pair is given by
\begin{eqnarray}
\Sigma_{\textrm{pg}}(P)=\sum_Qt_{\textrm{pg}}(Q)G_0(Q-P)
\end{eqnarray}
where $t_{\textrm{pg}}(Q)=\frac{-g'}{1-g'\chi(Q)}$ is the $t$-matrix due to non-condensed pairs. The pair susceptibility is constructed in the $G_0G$ formalism
$\chi(Q)=\sum_KG(K)G_0(Q-K)$, where $G$ is the full Green function with pairing fluctuation effect included. Here we assume that the coupling constant between fermions in non-condensed pairs, $g'$, is not necessarily equal to $g$, the coupling constant between fermions in condensed pairs. To determine the pairing onset temperature $T^*$, we still use the Thouless criteria, i.e., $t_{\textrm{pg}}(0)$ is divergent at $T^*$, or
\begin{eqnarray}
t^{-1}_{\textrm{pg}}(0)=1+g'\chi(0)=0.
\end{eqnarray}
One possible reason that $g'$ may not be equal to $g$ is that the Thouless criteria can not reduce to the BCS gap equation even when $g'=g$. Similarly, the $t$-matrix due to the condensed pair is $t_{\textrm{sc}}(Q)=-\frac{\Delta^2_{\textrm{sc}}}{T}\delta(Q)$, and the BCS self-energy is also expressed as $\Sigma_{\textrm{sc}}(P)=\sum_Qt_{\textrm{sc}}(Q)G_0(Q-P)$.
The order parameter is still determined by $\Delta_{\textrm{sc}}=-g\sum_{\mathbf{p}}\langle\psi_{\mathbf{p}\uparrow}\psi_{-\mathbf{p}\downarrow}\rangle$, which is non-zero below $T_c$. Now the full inverse Green function is given by
\begin{eqnarray}
G^{-1}(P)&=&G^{-1}_0(P)-\Sigma(P)\nonumber\\
&=&G^{-1}_0(P)-\Sigma_{\textrm{sc}}(P)-\Sigma_{\textrm{pg}}(P),
\end{eqnarray}
where $\Sigma(P)$ is the total self-energy. We emphasize that no further approximation is introduced now.

\subsection{Density Channel}
\begin{figure}[H]
\begin{center}
\includegraphics[width=0.7in,clip]{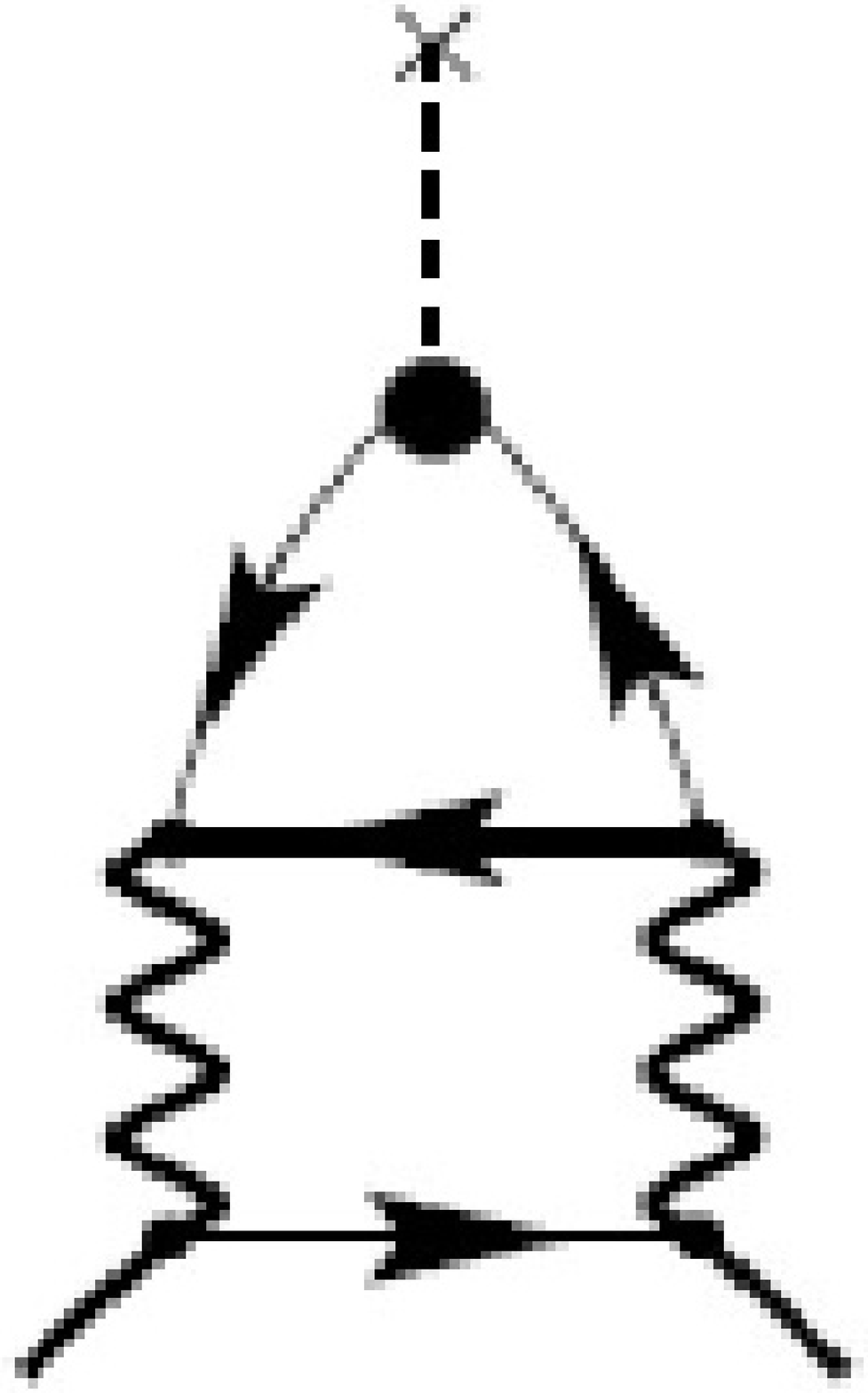}
\includegraphics[width=0.6in,clip]{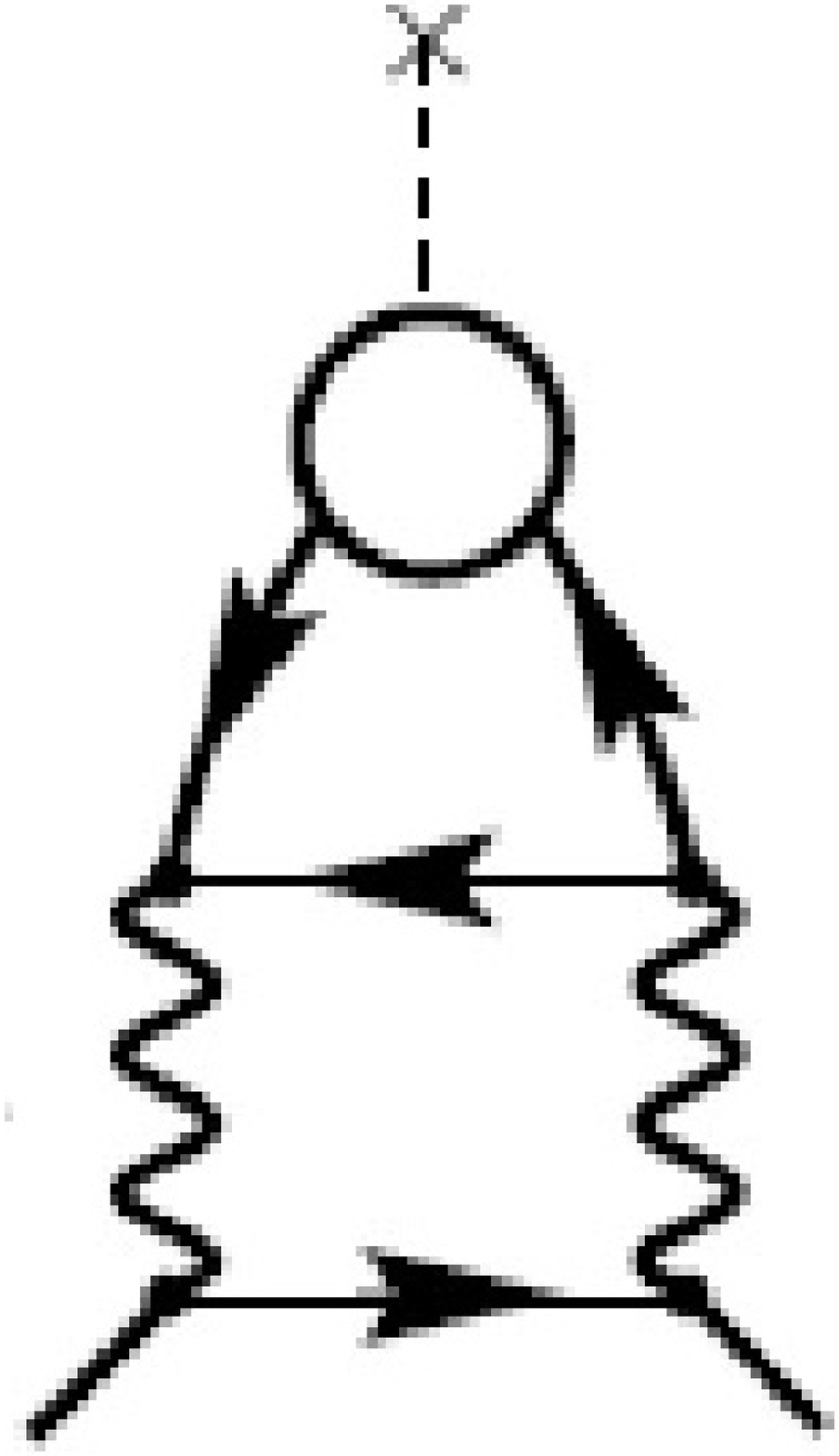}
\caption{The left one is the AL$_1$ diagram, and the right one is the AL$_2$ diagram. The thin and thick lines denote the bare and full Green's functions respectively. }
\label{fig:AL}
\end{center}
\end{figure}
In the density channel, to get a new gauge invariant interaction vertex, we must find a vertex correction which is consistent with the new self-energy $\Sigma_{\textrm{pg}}$. Such vertex does exist if we adapt the upper modified $G_0G$ formalism
\begin{eqnarray}\label{V1}
& &\Gamma^{\mu}(P+Q,P)=\gamma^{\mu}(P+Q,P)+\textrm{Coll}^{\mu}_{\textrm{sc}}(P+Q,P)\nonumber\\
&+&\textrm{MT}^{\mu}_{\textrm{sc}}(P+Q,P)+\textrm{MT}^{\mu}_{\textrm{pg}}(P+Q,P)\nonumber\\
&+&\textrm{AL}^{\mu}_{1}(P+Q,P)+\textrm{AL}^{\mu}_{2}(P+Q,P).
\end{eqnarray}
Here $\textrm{Coll}^{\mu}_{\textrm{sc}}$ and $\textrm{MT}^{\mu}_{\textrm{sc}}$ are still given by (\ref{C0}) and (\ref{MTSC}) respectively, and $\textrm{MT}^{\mu}_{\textrm{sc}}$ can be further expressed in a more general style by including the $t$-matrix $t_{\textrm{sc}}$
\begin{eqnarray}
& &\textrm{MT}^{\mu}_{\textrm{sc}}(P+Q,P)=\sum_Kt_{\textrm{sc}}(K)G_{0}(K-P)\times\nonumber\\
& &\gamma^{\mu}(K-P,K-P-Q)G_{0}(K-P-Q).
\end{eqnarray}
Hence the MT diagram for non-condensed pairs is given by
\begin{eqnarray}\label{MTmu}
& &\textrm{MT}^{\mu}_{\textrm{pg}}(P+Q,P)=\sum_Kt_{\textrm{pg}}(K)G_{0}(K-P)\times\nonumber\\
& &\gamma^{\mu}(K-P,K-P-Q)G_{0}(K-P-Q).
\end{eqnarray}
The fifth and sixth terms are two different types AL diagrams (shown in Figure.\ref{fig:AL}),
\begin{widetext}
\begin{eqnarray}\label{ALmu}
& &\textrm{AL}^{\mu}_{1}(P+Q,P)=-\sum_{K,L}t_{\textrm{pg}}(K)t_{\textrm{pg}}(K+Q)G_{0}(K-P)G(K-L)G_{0}(L+Q)\gamma^{\mu}(L+Q,L)G_{0}(L),\nonumber\\
& &\textrm{AL}^{\mu}_{2}(P+Q,P)=-\sum_{K,L}t_{\textrm{pg}}(K)t_{\textrm{pg}}(K+Q)G_{0}(K-P)G_{0}(K-L)G(L+Q)\Gamma^{\mu}(L+Q,L)G(L).
\end{eqnarray}
\end{widetext}
We see that $\textrm{AL}^{\mu}_{2}$ contains a full vertex, hence the expression (\ref{V1}) is in fact a series. In this scheme, the pseudogap effect does not enter into the terms related to the collective modes, this brings difficulties to the numerical work in the future. However, the theory is explicitly self-consistent because the $\textrm{Coll}^{\mu}_{\textrm{sc}}$ term vanishes above $T_c$ hence there are no Nambu-Goldstone modes exciations, which is consistent with the fact that the U(1) EM symmetry is unbroken.

The full vertex (\ref{V1}) satisfies the Ward identity
\begin{eqnarray}\label{WI2}
q_{\mu}\Gamma^{\mu}(P+Q,P)=G^{-1}(P+Q)-G^{-1}(P),
\end{eqnarray}
and the gauge invariant response functions now can be expressed as
\begin{eqnarray}\label{QmnG}
& &K^{\mu\nu}(Q)=\frac{n}{m}h^{\mu\nu}\\
&+&2\sum_{P}\Gamma^{\mu}(P+Q,P)G(P+Q) \gamma^{\nu}(P,P+Q)G(P).\nonumber
\end{eqnarray}
To prove the Ward identity we need a lemma
\begin{eqnarray}
& &q_{\mu}\big[\frac{1}{2}\textrm{AL}^{\mu}_{1}(P+Q,P)+\frac{1}{2}\textrm{AL}^{\mu}_{2}(P+Q,P)\nonumber\\
& &\quad+\textrm{MT}^{\mu}_{\textrm{pg}}(P+Q,P)\big]=0.
\end{eqnarray}
This proof of this lemma is outlined in the Appendix.\ref{app2}. Moreover, by applying the bare Ward identity (\ref{BWI}), we can show that
\begin{eqnarray}
& &q_{\mu}\textrm{MT}^{\mu}_{\textrm{pg}}(P+Q,P)\nonumber\\
&=&\sum_Kt_{\textrm{pg}}(K)\big[G(K-P-Q)-G(K-P)\big]\nonumber\\
&=&\Sigma_{\textrm{pg}}(P+Q)-\Sigma_{\textrm{pg}}(P).
\end{eqnarray}
Hence
\begin{eqnarray}\label{TWI}
& &q_{\mu}\big[\textrm{AL}^{\mu}_{1}(P+Q,P)+\textrm{AL}^{\mu}_{2}(P+Q,P)\nonumber\\
& &\quad+\textrm{MT}^{\mu}_{\textrm{pg}}(P+Q,P)\big]\nonumber\\
&=&\Sigma_{\textrm{pg}}(P)-\Sigma_{\textrm{pg}}(P+Q).
\end{eqnarray}
Finally, Eqs.(\ref{BWI}), (\ref{CWI}), (\ref{MWI}) and (\ref{TWI}) lead to the Ward identity (\ref{WI2}) self-consistently by assuming that $\textrm{AL}^{\mu}_{2}(P+Q,P)$ has a full gauge invariant interaction vertex.

Moreover, the full vertex should also respect the $Q$-limit Ward identity
\begin{eqnarray}\label{QWI2}
\lim_{\mathbf{q}\rightarrow\mathbf{0}}\Gamma^0(P+Q,P)|_{\omega=0}=\frac{\partial G^{-1}(P)}{\partial \mu}=1-\frac{\partial \Sigma(P)}{\partial \mu}.
\end{eqnarray}
A brief derivation shows that this identity ensures the compressibility sum rule
\begin{eqnarray}\label{QmnG3}
& &\frac{\partial n}{\partial \mu}=-2\sum_PG^2(P)\frac{\partial G^{-1}(P)}{\partial \mu}\nonumber\\
&=&-\lim_{\mathbf{q}\rightarrow\mathbf{0}}\sum_P\Gamma^0(P+Q,P)G(P+Q)\nonumber\\
& &\quad\times\gamma^0(P,P+Q)G(P)|_{\omega=0}\nonumber\\
&=&-\lim_{\mathbf{q}\rightarrow\mathbf{0}}K^{00}(Q)|_{\omega=0},
\end{eqnarray}
where $n=2\sum_PG(P)$ now. Note
\begin{eqnarray}\label{V20}
& &\Gamma^{\mu}(P+Q,P)=\Gamma^{\mu}_{\textrm{sc}}(P+Q,P)+\textrm{MT}^{\mu}_{\textrm{pg}}(P+Q,P)\nonumber\\
& &\quad+\textrm{AL}^{\mu}_{1}(P+Q,P)+\textrm{AL}^{\mu}_{2}(P+Q,P).
\end{eqnarray}
and that $\Gamma^{\mu}_{\textrm{sc}}(P+Q,P)$ already satisfies (\ref{QWI}), we only need to show that
\begin{eqnarray}\label{QWI3}
& &\lim_{\mathbf{q}\rightarrow\mathbf{0}}\big[\textrm{MT}^{0}_{\textrm{pg}}(P+Q,P)+\textrm{AL}^{0}_{1}(P+Q,P)\nonumber\\
&+&\textrm{AL}^{0}_{2}(P+Q,P)\big]|_{\omega=0}=-\frac{\partial \Sigma_{\textrm{pg}}(P)}{\partial \mu}.
\end{eqnarray}
It can be proved as following
\begin{eqnarray}\label{tmp1}
& &-\frac{\partial \Sigma_{\textrm{pg}}(P)}{\partial \mu}=-\sum_K t_{\textrm{pg}}(K)\frac{\partial G_0(K-P)}{\partial \mu}\nonumber\\
&-&\sum_K\frac{\partial t_{\textrm{pg}}(K)}{\partial \mu}G_0(K-P)\nonumber\\
&=&\sum_K t_{\textrm{pg}}(K)G^2_0(K-P)\frac{\partial G^{-1}_0(K-P)}{\partial \mu}\nonumber\\
&+&\sum_K t^2_{\textrm{pg}}(K)\frac{\partial t^{-1}_{\textrm{pg}}(K)}{\partial \mu}G_0(K-P)\nonumber\\
&=&\sum_K t_{\textrm{pg}}(K)G^2_0(K-P)\nonumber\\
&+&\sum_K t^2_{\textrm{pg}}(K)\frac{\partial \chi_{\textrm{pg}}(K)}{\partial \mu}G_0(K-P)\nonumber\\
&=&\sum_K t_{\textrm{pg}}(K)G^2_0(K-P)\nonumber\\
&+&\sum_{K,L}t^2_{\textrm{pg}}(K)\frac{\partial G_0(K-L)}{\partial \mu}G(L)G_0(K-P)\nonumber\\
&+&\sum_{K,L}t^2_{\textrm{pg}}(K)G_0(K-L)\frac{\partial G(L)}{\partial \mu}G_0(K-P)\nonumber\\
&=&\sum_K t_{\textrm{pg}}(K)G^2_0(K-P)\nonumber\\
&-&\sum_{K,L}t^2_{\textrm{pg}}(K)G_0^2(K-L)G(L)G_0(K-P)\nonumber\\
&-&\sum_{K,L}t^2_{\textrm{pg}}(K)G_0(K-L)G^2(L)\frac{\partial G^{-1}(L)}{\partial \mu}G_0(K-P).\nonumber\\
\end{eqnarray}
Since the 0-component of the bare vertex $\gamma^{\mu}$ is always 1, then right-hand-side of Eq.(\ref{tmp1}) is indeed
\begin{eqnarray}\label{QWI4}
\textrm{MT}^{0}_{\textrm{pg}}(P,P)+\textrm{AL}^{0}_{1}(P,P)+\textrm{AL}^{0}_{2}(P,P)
\end{eqnarray}
if we compare with the expressions (\ref{MTmu}) and (\ref{ALmu}). Therefore the $Q$-limit Ward identity is also satisfied.

\subsection{Spin Channel}
\begin{figure}[H]
\begin{center}
\includegraphics[width=3.3in,clip]
{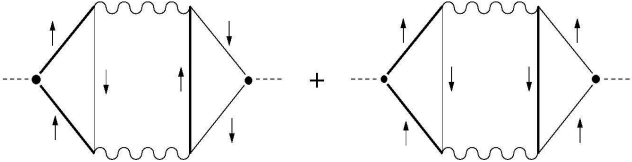}
\caption{Vertex corrections from spin-up AL$_1$ diagrams. }
\label{fig:S-V2}
\end{center}
\end{figure}
In the spin channel, the linear response theory is formulated in a similar but simpler way. The central idea is to find the gauge invariant spin interaction vertex. According to the expression of the spin interaction Hamiltonian (\ref{HIS0}), it is convenient to restore the pseudo-spin dependence of the Green function. Hence $G_{0\uparrow}(P)=G_{0\downarrow}(P)=G_0(P)$, $G_{\uparrow}(P)=G_{\downarrow}(P)=G(P)$. Generically, the vertex also contains the MT and AL diagrams when going beyond BCS theory
\begin{eqnarray}\label{VS1}
& &\Gamma^{\mu}_{\textrm{S}\sigma}(P+Q,P)=\gamma^{\mu}_{\textrm{S}\sigma}(P+Q,P)\nonumber\\
&+&\textrm{MT}^{\mu}_{\textrm{Ssc}\sigma}(P+Q,P)+\textrm{MT}^{\mu}_{\textrm{Spg}\sigma}(P+Q,P)\nonumber\\
&+&\textrm{AL}^{\mu}_{\textrm{S}1\sigma}(P+Q,P)+\textrm{AL}^{\mu}_{\textrm{S}2\sigma}(P+Q,P).
\end{eqnarray}
The MT diagram associated with the contributions from the order parameter and pseudogap are respectively given by
\begin{eqnarray}
& &\textrm{MT}^{\mu}_{\textrm{Ssc}\sigma}(P+Q,P)=\sum_Kt_{\textrm{sc}}(K)G_{0\bar{\sigma}}(K-P)\times\nonumber\\
& &\gamma^{\mu}_{\textrm{S}\bar{\sigma}}(K-P,K-P-Q)G_{0\bar{\sigma}}(K-P-Q),\nonumber\\
& &\textrm{MT}^{\mu}_{\textrm{Spg}\sigma}(P+Q,P)=\sum_Kt_{\textrm{pg}}(K)G_{0\bar{\sigma}}(K-P)\times\nonumber\\
& &\gamma^{\mu}_{\textrm{S}\bar{\sigma}}(K-P,K-P-Q)G_{0\bar{\sigma}}(K-P-Q).
\end{eqnarray}
Here we emphasize again that the spin interaction vertice have different signs for different pseudo-spin
indices, which leads to an important result that the contributions from the AL diagrams automatically cancel out. This can be shown by a straightforward verification. Figure.\ref{fig:S-V2} shows the vertex correction from the two spin-up AL$_1$ diagrams (with two sets of different pseudo-spin attributions). We have
\begin{eqnarray}
& &\textrm{AL}^{\mu}_{\textrm{S}1\uparrow}(P+Q,P)\nonumber\\
&=&-\sum_{K,K'}G_{0\downarrow}(K-P)t_{\textrm{pg}}(K+Q)t_{\textrm{pg}}(K')\times\nonumber\\
& &\big[G_{\uparrow}(K-K')G_{0\downarrow}(K'+Q)\gamma^{\mu}_{\textrm{S}\downarrow}(K'+Q,K')G_{0\downarrow}(K')\nonumber\\
&+&G_{\downarrow}(K-K')G_{0\uparrow}(K'+Q)\gamma^{\mu}_{\textrm{S}\uparrow}(K'+Q,K')G_{0\uparrow}(K')\big]\nonumber\\
&=&0,
\end{eqnarray}
where the fact that $G_{0\uparrow}=G_{0\downarrow}$, $G_{\uparrow}=G_{\downarrow}$ and $\gamma^{\mu}_{\textrm{S}\uparrow}=-\gamma^{\mu}_{\textrm{S}\downarrow}$ has been applied. Similar calculation indicates that the vertex corrections from the two spin-down AL$_1$ diagrams also vanish. The vertex corrections from spin-up AL$_2$ is
\begin{eqnarray}
& &\textrm{AL}^{\mu}_{\textrm{S}2\uparrow}(P+Q,P)\nonumber\\
&=&-\sum_{K,K'}G_{0\downarrow}(K-P)t_{\textrm{pg}}(K+Q)t_{\textrm{pg}}(K')\times\nonumber\\
& &\big[G_{0\uparrow}(K-K')G_{\downarrow}(K'+Q)\Gamma^{\mu}_{\textrm{S}\downarrow}(K'+Q,K')G_{\downarrow}(K')\nonumber\\
&+&G_{0\downarrow}(K-K')G_{\uparrow}(K'+Q)\Gamma^{\mu}_{\textrm{S}\uparrow}(K'+Q,K')G_{\uparrow}(K')\big]\nonumber\\
&=&0.
\end{eqnarray}
Hence the contributions from two spin-down AL$_2$ vanish too. From the equalities
\begin{eqnarray}
q_{\mu}\textrm{MT}^{\mu}_{\textrm{Ssc}\sigma}(P+Q,P)&=&S_{\sigma}\big[\Sigma_{\textrm{sc}}(P+Q)-\Sigma_{\textrm{sc}}(P)\big],\nonumber\\
q_{\mu}\textrm{MT}^{\mu}_{\textrm{Spg}\sigma}(P+Q,P)&=&S_{\sigma}\big[\Sigma_{\textrm{pg}}(P+Q)-\Sigma_{\textrm{pg}}(P)\big]\nonumber
\end{eqnarray}
one can show that the Ward identity for the full spin interaction vertex is satisfied
\begin{eqnarray}\label{SWI1}
q_{\mu}\Gamma^{\mu}_{\textrm{S}\sigma}(P+Q,P)=S_{\sigma}\big[G^{-1}(P+Q)-G^{-1}(P)\big].
\end{eqnarray}
The spin linear response theory is gauge invariant too.

Therefore all consistency constraints are satisfied within this scheme both in the density and spin channels when pairing fluctuation effects are considered. However, the approach in the density channel is not a useful form for numerical application. If any approximation is applied, it may most possibly violate some of the constraints. In certain situation, some constraints may survive, hence these consistency conditions can play as an indicator to ``measure'' how good the approximation is.

\section{Conclusion}
We have constructed gauge invariant density and spin linear response theories for a Fermi gas undergoing BCS-BEC crossover by including adequate diagrams in the interaction vertices using the $t$-matrix formalism based on a slightly corrected $G_0G$ scheme. We verified that the Ward identities and $Q$-limit Ward identity is satisfied when the contributions due to the order parameter (condensed pairs) and pseudogap (non-condensed pairs) are both included. This justifies Nambu's assertion that the modification of the vertex must be consistent with the way that the self-energy is included in the quasi-particle. Those constraints guarantee the self-consistency of the theories. Until now our approach is a purely theoretical formalism without including any approximation, yet we believe it will shed light on the reliable theoretical predictions of the transport properties of strongly correlated Fermi gases and help us to understand more about the many particle theory. Future improvements include trustworthy numerical calculations by taking suitable approximations.

Hao Guo thanks the support by National Natural Science Foundation of China (Grants No. 11204032) and Natural Science Foundation of Jiangsu Province, China (SBK201241926).

\appendix
\section{Vertex Correction in BCS mean field theory}\label{app1}
\begin{eqnarray}\label{tmp4}
\Pi^{\mu}&=&\frac{Q_3Q^{\mu}_4-Q_2Q^{\mu}_5}{Q_1Q_2-Q^3_3},\nonumber\\
\bar{\Pi}^{\mu}&=&\frac{Q_3Q^{\mu}_5-Q_1Q^{\mu}_4}{Q_1Q_2-Q^3_3}.
\end{eqnarray}
Here
\begin{eqnarray}\label{tmp5}
Q_1(Q)&=&\frac{1}{g}+\sum_PG_{\textrm{sc}}(P-Q)G_{\textrm{sc}}(-P),\nonumber\\
Q_2(Q)&=&\frac{1}{g}+\sum_PG_{\textrm{sc}}(P+Q)G_{\textrm{sc}}(-P),\nonumber\\
Q_3(Q)&=&-\sum_PF_{\textrm{sc}}(P+Q)F_{\textrm{sc}}(P),\nonumber\\
Q^{\mu}_4(Q)&=&-2\sum_P\gamma^{\mu}(P+Q,P)G_{\textrm{sc}}(P+Q)F_{\textrm{sc}}(P),\nonumber\\
Q^{\mu}_5(Q)&=&-2\sum_P\gamma^{\mu}(P+Q,P)F_{\textrm{sc}}(P+Q)G_{\textrm{sc}}(P).\nonumber\\
\end{eqnarray}

\section{Proof of the lemma}\label{app2}
Since $\Sigma_{\textrm{pg}}(P+Q)=\sum_Kt_{\textrm{pg}}(K)G_0(K-P-Q)=\sum_Kt_{\textrm{pg}}(K+Q)G_0(K-P)$, therefore

\begin{eqnarray}\label{eqn:ac}
& &0=\sum_K\big[t_{\textrm{pg}}(K+Q)G_0(K-P) \nonumber\\
& &\quad\qquad -t_{\textrm{pg}}(K)G_0(K-P-Q)\big] \nonumber\\
&=&\sum_K\Big(\big[t_{\textrm{pg}}(K+Q)-t_{\textrm{pg}}(K)\big]G_0(K-P)\nonumber\\
&+&t_{\textrm{pg}}(K)[G_0(K-P)-G_0(K-P-Q)\big]\Big)\nonumber\\
&=&\sum_K\Big(t_{\textrm{pg}}(K)[G_0(K-P)-G_0(K-P-Q)\big]\nonumber\\
&-&t_{\textrm{pg}}(K+Q)t_{\textrm{pg}}(K)\big[\chi(K+Q)-\chi(K)\big]G_0(K-P)\Big).\nonumber
\end{eqnarray}
Note $\chi(K)=\sum_{L}G(K-L)G_0(L)=\sum_{L}G_0(K-L)G(L)$, we have
\begin{eqnarray}
& &\chi(K+Q)-\chi(K)\nonumber\\
&=&\frac{1}{2}\sum_{L}\Big(G(K-L)\big[G_0(L+Q)-G_0(L)\big]\nonumber\\
& &\qquad+G_0(K-L)\big[G(L+Q)-G(L)\big]\Big).
\end{eqnarray}
Using this equality, we get
\begin{eqnarray}\label{eqn:ac2}
& &0=-\frac{1}{2}\sum_{KL}t_{\textrm{pg}}(K+Q)t_{\textrm{pg}}(K)G_0(K-P)\nonumber\\
&\times&\Big(G(K-P-Q)\big[G_0(L+Q)-G_0(L)\big]\nonumber\\
&+&G_0(K-P-Q)\big[G(L+Q)-G(L)\big]\Big)\nonumber\\
&+&\sum_Kt_{\textrm{pg}}(K)[G_0(K-P)-G_0(K-P-Q)\big].
\end{eqnarray}
By applying the bare WI (\ref{BWI}) and WI (\ref{WI2}), we can see that Eq.(\ref{eqn:ac2}) leads to the lemma.


\end{document}